\documentclass{article}
\usepackage{geometry}[margin=1cm]
\usepackage{graphicx}          
\usepackage{tikz}
\usetikzlibrary{arrows.meta}   
\usepackage{natbib}
\definecolor{inkP}{HTML}{0B0B0B}
\definecolor{inkS}{HTML}{52514E}
\definecolor{gridc}{HTML}{E1E0D9}
\definecolor{neutral}{HTML}{F0EFEC}
\definecolor{rampA}{HTML}{86B6EF}
\definecolor{rampB}{HTML}{2A78D6}
\definecolor{rampC}{HTML}{104281}

\usepackage{times}

\usepackage{amsmath,amsfonts,bm}

\def\eqref#1{equation~\ref{#1}}

\def\1{\bm{1}}

\DeclareMathAlphabet{\mathsfit}{\encodingdefault}{\sfdefault}{m}{sl}
\SetMathAlphabet{\mathsfit}{bold}{\encodingdefault}{\sfdefault}{bx}{n}

\definecolor{cInk}{RGB}{45,45,45}\definecolor{cGrey}{RGB}{125,125,125}
\definecolor{cBlue}{RGB}{58,101,166}\definecolor{cBlueF}{RGB}{225,235,250}
\definecolor{cAmb}{RGB}{200,118,38}\definecolor{cAmbF}{RGB}{252,238,214}
\definecolor{cGrn}{RGB}{48,138,82}\definecolor{cGrnF}{RGB}{222,242,228}
\definecolor{cRed}{RGB}{186,72,72}\definecolor{cRedF}{RGB}{250,228,228}
\definecolor{cSand}{RGB}{242,226,190}\definecolor{cSandL}{RGB}{190,165,110}
\definecolor{cSea}{RGB}{226,238,247}\definecolor{cPaper}{RGB}{255,253,247}
\definecolor{cGreyF}{RGB}{232,232,232}

\usepackage[utf8]{inputenc}
\usepackage[T1]{fontenc}
\usepackage{hyperref}
\usepackage{url}
\usepackage{booktabs}
\usepackage{amsmath,amssymb,amsthm}
\usepackage{xcolor}
\usepackage{graphicx}
\usepackage{multirow}
\usepackage{enumitem}
\usepackage{microtype}
\usepackage{array}
\usepackage{float}
\usepackage{algorithm}
\usepackage{algpseudocode}
\usepackage{tikz}
\usetikzlibrary{arrows.meta,positioning,fit,calc}
\usepackage{pgfplots}
\pgfplotsset{compat=1.18}

\newcommand{\sys}{\textsc{LinCodeEvolve}}

\newcommand{\Ftwo}{\mathbb{F}_2}
\newcommand{\wt}{\operatorname{wt}}

\newcommand{\cnk}[2]{[#1,#2]}

\theoremstyle{plain}

\theoremstyle{definition}

\theoremstyle{remark}

\title{Evolving Towards Better Codes: LLM-Guided Search for High-Distance Binary Linear Codes}

\author{
Amal Seddas\textsuperscript{1},
Vladyslav Shashkov\textsuperscript{1},
Maryna Viazovska\textsuperscript{1},
Emmanuel Abbe\textsuperscript{1}\\
\textsuperscript{1}EPFL
}
\usetikzlibrary{arrows.meta, calc, shadows}
\date{}

\begin{document}

\maketitle
\begin{abstract}
Evolutionary program search driven by large language models (LLMs) has produced
record-breaking constructions for open problems in combinatorics and beyond. We apply this approach to the longstanding problem of improving the best-known bounds for binary linear codes. Building on the EvoTune evolutionary
framework and the ShinkaEvolve codebase, we introduce \sys{}, which evolves
code-construction programs against an exact minimum-distance
evaluator. A strategy loop combines diversity-driven search and expert supervision:
when progress plateaus, new strategies are used to redirect the search.
\sys{} discovers seven record-breaking codes, $[172,21,66]$, $[173,20,68]$,
$[176,21,68]$, $[181,21,70]$, $[184,21,72]$, $[189,22,72]$ and $[200,21,77]$, six
of which have concise quasi-cyclic descriptions. With standard code modification techniques,
they improve $22$ entries of the tables. Every code is verified by
exhaustive enumeration; the results have been checked by the maintainer of the
code tables and will be incorporated into them. These results suggest that LLM-guided search can help find improved codes and complement existing methods in coding theory.
\end{abstract}

\section{Introduction}

We study the construction of binary linear codes with large minimum distance. This is a central problem in coding theory: minimum distance determines a code's worst-case error-correction capability and enters standard upper bounds on block error probability under maximum-likelihood decoding on the binary symmetric channel \citep[Eq.~(2.4), p.~15]{sason2006performance}. The Gilbert-Varshamov bound establishes the existence of codes with favorable parameters \citep{varshamov1957estimate}, yet explicit constructions can exceed this bound for particular lengths and dimensions, as recorded in \href{https://www.codetables.de}{CodeTables}.\footnote{A database of best known lower and upper bounds on the minimum distance of linear codes.} Finding such constructions improves known lower bounds and can reveal structure shared by codes with large minimum distance.

A construction that works well at one length and dimension need not work equally well at another. Progress therefore calls for selecting, adapting, and combining algebraic constructions and computational search procedures across parameter pairs. This makes code construction a natural setting for automated program discovery: a proposed construction can be executed, its output evaluated, and the result used to guide the next proposal.

FunSearch \citep{romera2024funsearch} implements this idea by pairing language-model-generated programs with an executable evaluator and a database of earlier candidates. High-scoring programs inform subsequent proposals, allowing the search to refine promising ideas while exploring alternatives. FunSearch produced new cap-set constructions and bin-packing heuristics. AlphaEvolve \citep{novikov2025alphaevolve} extends program evolution from individual functions to codebases with multiple evolved components. Georgiev et al.\ \citep{georgiev2025} combine this approach with reasoning and proof-assistant systems for mathematical discovery and, in some cases, automated proof. In coding theory, Weindel and Heckel \citep{weindel2025llm} use FunSearch to evolve priority functions for greedy constructions of deletion-correcting codes, recovering the conjectured-optimal Varshamov-Tenengolts construction for a single deletion and improving lower bounds for multiple deletions and quaternary edit codes.

We introduce \sys, a program-evolution framework for discovering binary linear codes with improved parameters. Building on ShinkaEvolve \citep{lange2025shinka}, \sys coordinates agents that propose and implement experiments with specialist search islands focused on different construction strategies. They draw on a shared bank of strategies and verified codes, initialized with expert-written implementations of diverse coding-theoretic constructions. An executable evaluator assesses proposed codes, and verified results guide later proposals. This separation between proposing a construction and evaluating its output is especially important here: determining minimum distance is computationally difficult in general, while improvements to the best-known bound occur in discrete steps.

Our contributions are a framework for systematically evolving linear-code constructions and new codes that improve longstanding bounds. Each search run targets a fixed length dimension pair $(n,k)$. Within the range $1 \leq k \leq n \leq 200$, \sys discovers seven parent codes with parameters $[173,20,68]$, $[172,21,66]$, $[176,21,68]$, $[181,21,70]$, $[184,21,72]$, $[189,22,72]$, and $[200,21,77]$. Each improves a longstanding lower bound recorded in \href{https://www.codetables.de}{CodeTables} by one to three units. Explicit derivations from these parent codes yield improvements for $22$ parameter sets at lengths 171-200 (Section~\ref{New Records}).

\section{Background and Related Work}
\label{sec:background}

\paragraph{Binary linear codes}
Consider $k,n \in \mathbb{N},\, k \leq n$. A binary linear $\cnk{n}{k}$ code $\mathcal{C}$ is a $k$-dimensional subspace of $\Ftwo^n$. We will refer to $R:=\frac{k}{n}$ as the code rate. A full-rank generator matrix $G \in \Ftwo^{k \times n}$ maps a message vector $u \in \Ftwo^k$ to a codeword $c = uG \in \mathcal{C}$. By linearity, the minimum distance $d(\mathcal{C})$ is equivalent to the minimum Hamming weight among all nonzero codewords in $\mathcal{C}$:
\[
    d(\mathcal{C}) := \min_{u \in \Ftwo^k \setminus \{0\}} \wt(uG)=\min_{u \in \Ftwo^k \setminus \{0\}}\left|\left\{ i \in [n] \big{|} \left(uG\right)_i = 1\right\}\right|.
\]
The minimum distance establishes the bounded-distance decoding capability of the code, guaranteeing the correction of up to $\lfloor (d(\mathcal{C})-1)/2 \rfloor$ arbitrary errors. The multiplicity of weight-$w$ codewords is denoted by $A_w$, which defines the weight enumerator of $\mathcal{C}$, i.e. the vector $\left\{A_w|w \in [n]\right\}$.
\paragraph{ Gilbert-Varshamov bound} Consider a parameter $\delta \in \left[0, \frac{1}{2}\right)$. Define the binary entropy function $h_2(x)=-x\log_2(x)-(1-x)\log_2(1-x)$. Then, for any $\epsilon>0$ and all sufficiently large $n$, there exists a binary linear code $\mathcal{C}$ of length $n$, dimension $k$, and minimum distance $d(\mathcal{C})$ satisfying
\begin{itemize}
    \item $k \geq n(1-h_2(\delta)-\epsilon)$
    \item $d(\mathcal{C}) \geq n\delta$
\end{itemize}
At a high level, the Gilbert-Varshamov bound provides an achievability guarantee for the tradeoff between the code rate and relative minimum distance: asymptotically, binary linear codes with relative minimum distance at least $\delta$ exist at any rate below $1-h_2(\delta)$.

\paragraph{Record tables and distance certification.}
For fixed $(n,k)$, the construction problem is to find a binary
linear code with the largest possible minimum distance.
Grassl's tables \citep{grassl2006searching}\footnote{\url{http://www.codetables.de}}
record the best-known lower and upper bounds on this optimum,
together with supporting constructions and references.
A construction with certified distance at least $d$ establishes a
lower bound; upper bounds limit what any code with those parameters
can achieve. These include the Griesmer bound
\citep{griesmer1960bound} and linear-programming bounds.

Certifying a candidate is computationally challenging:
computing minimum distance is NP-hard \citep{vardy1997intractability}.
Exact methods include the Brouwer-Zimmermann algorithm,
implemented in Magma and Sagemath \citep{bosma1997magma,grassl2006searching},
and, for small $k$, enumeration of all $2^k-1$ nonzero messages.
Randomized searches based on information-set decoding
\citep{prange1962information} can find low-weight codewords and
therefore upper bounds on a candidate's minimum distance.
However, failing to find a low-weight codeword does not certify
its absence. Improving a table's lower bound consequently requires
a rigorous distance guarantee.

\paragraph{Computer search for record codes.}
Existing searches exploit algebraic structure to make the space of
candidate codes more manageable.
One major approach searches within quasi-cyclic and quasi-twisted
families. This line includes early computational tables
\citep{chen1969quasi,gulliver1991some}, structural results for
one-generator codes \citep{aydin2001structure}, and subsequent
search algorithms \citep{daskalov2003new,chen2015iterative}.
A second approach uses \emph{secondary constructions} to obtain
new codes from known ones. Examples include Construction X and XX
\citep{sloane1972new}, shortening, puncturing, extension, and
subcodes \citep{huffman2003fundamentals}, as well as augmentation
methods that add generator rows to a quasi-cyclic parent
\citep{chen2015augmentation,aydin2017augmentation}.
A third approach prescribes a symmetry group and searches for
codes invariant under its action, for example by solving
Diophantine systems \citep{betten2006error} or constructing group
codes over semidirect products \citep{yu2024group}.
Together, these approaches motivate searching across construction
families, parent codes, and symmetries while retaining rigorous
distance verification.
Learning-based methods have also been explored for specialized
families, including reinforcement learning for linear
complementary-dual codes \citep{wu2025lcd}.
Our focus is the search for improvements to the general binary
linear-code tables.

\paragraph{LLM-driven program evolution.}
Evolution through Large Models \citep{lehman2022elm} and Language Model
Crossover \citep{meyerson2023lmx} use an LLM to mutate and recombine
programs. FunSearch \citep{romera2024funsearch}, AlphaEvolve
\citep{novikov2025alphaevolve} and ShinkaEvolve \citep{lange2025shinka}
apply the same loop to open mathematical and algorithmic problems, EvoTune \citep{evotune} guides the mutation through Direct Preference Optimization (DPO) given the scores assigned to the programs by the separately-defined reward function;
\citep{wu2024evolutionary} survey the field. \sys{} is built on
ShinkaEvolve and keeps its loop: sample a parent and a few inspiration
programs from an archive split into islands, ask an LLM for a patch or a
new program, evaluate it, store it. It also keeps the novelty judge, the
meta summarizer, prompt evolution and UCB1-style model selection
\citep{auer2002finite}. Two things change. The single proposal call
becomes a chain of oracle, strategist, contract repair and implementer,
with a helper-free invent operator as control. And nothing enters the
archive without an exact certificate (Section~\ref{sec:framework}).

\paragraph{Diversity and verified feedback.}
In our runs the model often handed back the parent matrix unchanged. This
matches \citep{gurkan2026mutation}, who show that LLM mutation chains
collapse onto a few forms even without selection pressure.
Quality-diversity archives \citep{mouret2015illuminating} and island
models \citep{alba2002parallelism} are the basis of our family-balanced
archive and our islands. Learned constructions
\citep{wagner2021constructions,charton2024patternboost} and
neuro-symbolic provers \citep{trinh2024solving} share our premise that a
learned proposer needs an exact checker. Automated research agents
\citep{lu2024aiscientist,hu2024automated} inspired the split of work
among our roles. In \sys{} only exact computations reach a score or the
stored evidence; what a program says about itself never changes its
score.
\section{Problem Statement}
\label{sec:problem}
Fix parameters $(n,k)$ where the table entry is open, meaning the lower bound $d_{\mathrm{LB}}$ is strictly less than the upper bound $d_{\mathrm{UB}}$. The task is to find the generator matrix $G$ which generates the binary linear code $\mathcal{C}$ with $d(\mathcal{C}) \ge d_{\mathrm{LB}} + 1$, meaning we beat the best-known linear code (\textsc{BKLC}) for these parameters. 

To do this, we track two values: the minimum distance $d$, and the number of minimum-weight codewords $A_d$. The latter is called the \emph{shell}. Between two codes with the same minimum distance, we prefer the one with the smaller shell. To reach $d+1$, the amount of weight-$\leq d$ codewords must be $0$, i.e. the shell must be equal to $0$, making $A_d$ a useful metric. Minimizing the shell given that the minimum distance is $d$ guides the search in the direction of better codes.

\section{The \sys{} Framework}
\label{sec:framework}
\sys{} evolves construction programs to improve best-known lower bounds
on the minimum distance of binary linear codes. It follows the
evolutionary search loop of EvoTune \citep{evotune}: a
language model proposes programs, their outputs are evaluated, and
archived results guide subsequent proposals. ShinkaEvolve
\citep{lange2025shinka} provides the execution infrastructure and the
program database, including islands and migration, novelty assessment
and prompt evolution. We retain these components. The
code-construction components summarised in Table~\ref{tab:records} are
built on top of them, and the language model's parameters remain fixed
throughout the search.

Three properties of code construction motivate these additions. First,
minimum distance is integer-valued and gives coarse feedback: many
candidates share the same distance, so we also compare their numbers of
minimum-weight codewords. Second, returning the parent code produces a
valid output without advancing the search, so the loop must recognise
copies. Third, progress can require a change of construction family,
so algebraic tools and evidence about routes already explored must be
available when the next proposal is written. At the dimensions we
consider, an exact evaluation of one code takes seconds while a model
proposal takes minutes, so each generated program should search many
candidates during one execution. \sys{} organises this search around
five components:
\begin{itemize}\setlength{\itemsep}{2pt}\setlength{\parskip}{0pt}\setlength{\parsep}{0pt}
  \item \textbf{An agent team.}
    The \emph{discover} operator separates proposal generation into
    three roles: an oracle analyses the current code, a strategist
    compares construction approaches and commits to one in a
    structured plan, and an implementer writes the corresponding
    program. A separate \emph{invent} operator generates programs
    without direct access to the Toolkit.
  \item \textbf{A Toolkit.}
    Generated programs can call hand-written strategies that implement
    constructions and search procedures from the coding-theory
    literature, so an expert construction costs a few lines rather than
    a new implementation, and proposals can select, combine and extend
    them.
  \item \textbf{Evaluation and search feedback.}
    The evaluator computes the full weight distribution of each
    admissible returned code. The fitness is the lexicographic order on
    $(d,-A_d)$: higher minimum distance first, then fewer minimum-weight
    codewords; it is used for champion selection and parent ranking
    alike.
  \item \textbf{Specialised islands and construction-based
    initialisation.}
    Each island's prompts carry one construction family, such as
    algebraic codes over extension fields or sparse-graph parity-check
    design, to encourage exploration of different construction
    families. The initial code is selected from a portfolio of
    constructions after exact evaluation of the shortlisted candidates.
  \item \textbf{Exact verification and human input.}
    Only evaluator-computed results determine a code's fitness; a
    program's self-reported claims are stored but cannot change it.
    Human notes, tools and certificates enter through shared files that
    the prompts read, and interventions are made through explicit
    pause-and-resume controls. 
\end{itemize}
The model writes the search program: it chooses a construction
approach, decides whether to modify the parent or start again, and
specifies how candidates are selected. The framework chooses parents
and proposal operators, runs the programs, evaluates their outputs,
and updates the archive, retaining one program per distinct row space.
Figure~\ref{fig:framework} shows one generation; the following
subsections describe its components.

\definecolor{cInk}{RGB}{45,45,45}\definecolor{cGrey}{RGB}{125,125,125}
\definecolor{cBlue}{RGB}{58,101,166}\definecolor{cBlueF}{RGB}{225,235,250}
\definecolor{cAmb}{RGB}{200,118,38}\definecolor{cAmbF}{RGB}{252,238,214}
\definecolor{cGrn}{RGB}{48,138,82}\definecolor{cGrnF}{RGB}{222,242,228}
\definecolor{cRed}{RGB}{186,72,72}\definecolor{cRedF}{RGB}{250,228,228}
\definecolor{cSand}{RGB}{242,226,190}\definecolor{cSandL}{RGB}{190,165,110}
\definecolor{cSea}{RGB}{226,238,247}\definecolor{cPaper}{RGB}{255,253,247}
\definecolor{cGreyF}{RGB}{232,232,232}
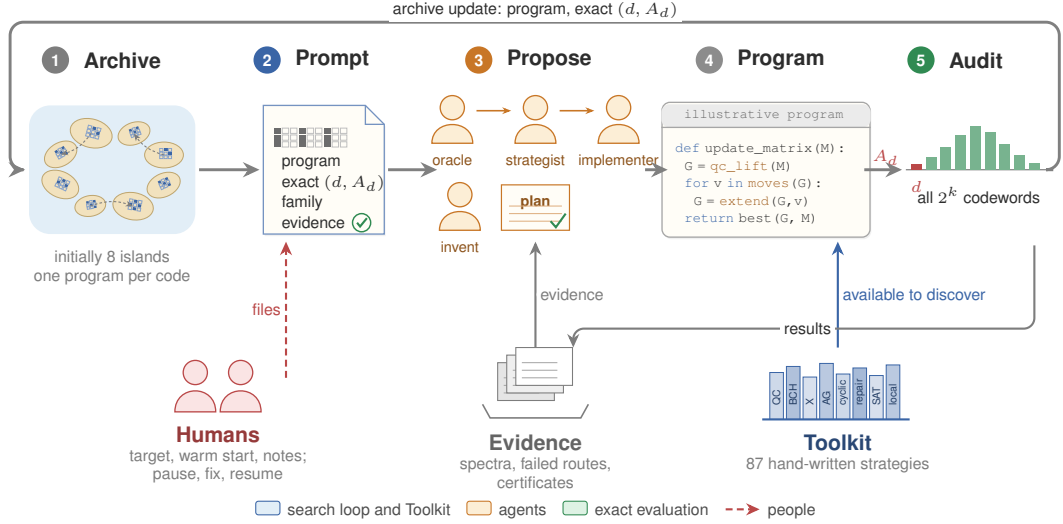
\begin{figure}[h]
\centering
\begin{tikzpicture}[x=1cm,y=1cm, font=\sffamily, text=cInk,
  flow/.style={-{Stealth[length=2.2mm,width=2mm]}, line width=1.1pt, black!50, rounded corners=5pt},
  feed/.style={-{Stealth[length=1.8mm,width=1.6mm]}, line width=0.8pt, cBlue, rounded corners=4pt},
  evd/.style={-{Stealth[length=1.8mm,width=1.6mm]}, line width=0.8pt, cGrey, rounded corners=4pt},
  hum/.style={-{Stealth[length=1.8mm,width=1.6mm]}, line width=0.8pt, cRed, dash pattern=on 2.6pt off 1.6pt, rounded corners=4pt},
  lab/.style={font=\sffamily\fontsize{6}{7}\selectfont, fill=white, inner sep=1.5pt, align=center},
  body/.style={font=\sffamily\fontsize{6}{7}\selectfont, align=center},
  ttl/.style={font=\sffamily\bfseries\footnotesize, align=center},
  num/.style={circle, fill=#1, text=white, font=\sffamily\bfseries\fontsize{6}{6}\selectfont, inner sep=0.5pt, minimum size=3.6mm},
  tag/.style={font=\sffamily\fontsize{5.5}{6}\selectfont, rounded corners=1.5pt, inner sep=1.8pt, text=#1!60!black, fill=#1, draw=#1!60!black, line width=0.35pt}
]
\def\xa{0}\def\xb{2.8}\def\xc{5.6}\def\xd{8.65}\def\xe{11.45}
\begin{scope}[shift={(\xa,0.9)}]
  \fill[cSea, rounded corners=8pt] (-1.1,-0.85) rectangle (1.1,0.85);
  \foreach \i/\r/\s in {0/0/1.0,1/20/0.85,2/-15/1.1,3/30/0.9,4/-25/1.0,5/10/0.8,6/-10/1.05,7/25/0.9}{
    \pgfmathsetmacro{\a}{\i*45+22}
    \begin{scope}[shift={({0.74*cos(\a)},{0.52*sin(\a)})}, rotate=\r, scale=\s*0.95]
      \fill[cSand, draw=cSandL, line width=0.35pt] (0,0) ellipse (0.27 and 0.18);
      \foreach \x in {-1,0,1}{\foreach \y in {-1,0,1}{\pgfmathparse{int(mod(\x*3+\y*5+\i,2))}
        \ifnum\pgfmathresult=1 \fill[cBlue] (\x*0.06,\y*0.05) rectangle ++(0.045,0.04);\else \draw[cBlue!60,line width=0.2pt] (\x*0.06,\y*0.05) rectangle ++(0.045,0.04);\fi}}
    \end{scope}}
  \foreach \i in {0,2,5}{\pgfmathsetmacro{\a}{\i*45+22}\pgfmathsetmacro{\b}{\i*45+67}
    \draw[-{Stealth[length=1mm]}, cGrey, line width=0.4pt, dash pattern=on 1pt off 0.8pt] ({0.74*cos(\a)},{0.52*sin(\a)}) to[bend left=20] ({0.74*cos(\b)},{0.52*sin(\b)});}
\end{scope}
\node[body, text=cGrey, anchor=north] at (\xa,-0.05) {initially 8 islands\\ one program per code};
\begin{scope}[shift={(\xb,0.9)}]
  \fill[cPaper, draw=cBlue!70, line width=0.5pt, drop shadow={opacity=0.12, shadow xshift=0.5pt, shadow yshift=-0.8pt}]
    (-0.8,-0.85) -- (-0.8,0.85) -- (0.55,0.85) -- (0.8,0.6) -- (0.8,-0.85) -- cycle;
  \draw[cBlue!70, line width=0.5pt] (0.55,0.85) -- (0.55,0.6) -- (0.8,0.6);
  \foreach \x in {0,...,8}{\foreach \y in {0,...,2}{\pgfmathparse{int(mod(\x*7+\y*3,3))}
    \ifnum\pgfmathresult=0 \fill[cInk!70] (-0.66+\x*0.11,0.3+\y*0.09) rectangle ++(0.085,0.07);\else \draw[cInk!30,line width=0.2pt] (-0.66+\x*0.11,0.3+\y*0.09) rectangle ++(0.085,0.07);\fi}}
  \node[body, anchor=west, align=left] at (-0.68,0.08) {program};
  \node[body, anchor=west, align=left] at (-0.68,-0.18) {exact $(d,A_d)$};
  \node[body, anchor=west, align=left] at (-0.68,-0.44) {family};
  \node[body, anchor=west, align=left] at (-0.68,-0.7) {evidence};
  \draw[cGrn, line width=0.55pt] (0.5,-0.7) circle (0.12); \draw[cGrn, line width=0.8pt] (0.44,-0.7)--(0.48,-0.75)--(0.57,-0.64);
\end{scope}
\newcommand{\avatar}[2]{\begin{scope}[shift={#1}]
  \fill[cAmbF, draw=cAmb, line width=0.5pt] (0,0.17) circle (0.16);
  \fill[cAmbF, draw=cAmb, line width=0.5pt] (-0.27,-0.3) .. controls (-0.27,0.03) and (0.27,0.03) .. (0.27,-0.3) -- cycle;
  \node[font=\sffamily\fontsize{5.5}{6}\selectfont, anchor=north, text=cAmb!70!black] at (0,-0.33) {#2};\end{scope}}
\begin{scope}[shift={(\xc,0.9)}]
  \avatar{(-1.1,0.65)}{oracle}
  \avatar{(0,0.65)}{strategist}
  \avatar{(1.1,0.65)}{implementer}
  \draw[-{Stealth[length=1.2mm]}, cAmb, line width=0.6pt] (-0.78,0.78) -- (-0.32,0.78);
  \draw[-{Stealth[length=1.2mm]}, cAmb, line width=0.6pt] (0.32,0.78) -- (0.78,0.78);
  \fill[cPaper, draw=cAmb, line width=0.45pt] (-0.45,-0.8) rectangle (0.45,-0.2);
  \node[font=\sffamily\bfseries\fontsize{5.5}{6}\selectfont, text=cAmb!70!black, anchor=north] at (0,-0.22) {plan};
  \foreach \y in {-0.48,-0.58,-0.68}{\draw[cAmb!60, line width=0.35pt] (-0.35,\y)--(0.35,\y);}
  \draw[cGrn, line width=0.8pt] (0.2,-0.68)--(0.27,-0.76)--(0.4,-0.58);
  \avatar{(-1.0,-0.5)}{invent}
\end{scope}
\node[draw=black!35, fill=cPaper, rounded corners=3pt, line width=0.5pt, inner sep=5pt, text width=2.4cm, align=left,
      font=\ttfamily\fontsize{5.2}{6.6}\selectfont, drop shadow={opacity=0.12, shadow xshift=0.5pt, shadow yshift=-0.8pt}] (prog) at (\xd,0.9)
  {\rule{0pt}{0.25cm}\\ \textcolor{cBlue}{def} update\_matrix(M):\\
   \ \ G = \textcolor{cAmb!80!black}{qc\_lift}(M)\\
   \ \ \textcolor{cBlue}{for} v \textcolor{cBlue}{in} \textcolor{cAmb!80!black}{moves}(G):\\
   \ \ \ \ G = \textcolor{cAmb!80!black}{extend}(G,v)\\
   \ \ \textcolor{cBlue}{return} best(G, M)};
\fill[black!8, rounded corners=3pt] ([yshift=-1pt]prog.north west) rectangle ([yshift=-0.34cm]prog.north east);
\draw[black!35, line width=0.5pt] ([yshift=-0.34cm]prog.north west) -- ([yshift=-0.34cm]prog.north east);
\node[font=\ttfamily\fontsize{5}{5}\selectfont, text=black!60, anchor=west] at ([xshift=5pt,yshift=-0.17cm]prog.north west) {illustrative program};
\begin{scope}[shift={(\xe,0.9)}]
  \draw[cGrey, line width=0.45pt] (-0.95,0.0) -- (0.95,0.0);
  \foreach \i/\h in {0/0.07,1/0.16,2/0.3,3/0.46,4/0.58,5/0.5,6/0.34,7/0.19,8/0.09}{\fill[cGrn!65] (-0.88+\i*0.195,0.0) rectangle ++(0.14,\h);}
  \fill[cRed] (-0.88,0.0) rectangle ++(0.14,0.07);
  \node[body, text=cRed, anchor=north] at (-0.81,-0.02) {$d$};
  \node[body, text=cRed, anchor=east] at (-0.92,0.16) {$A_d$};
  \node[body, anchor=north east] at (0.95,-0.1) {all $2^k$ codewords};
\end{scope}
\foreach \x/\c/\i/\t in {\xa/cGrey/1/Archive,\xb/cBlue/2/Prompt,\xc/cAmb/3/Propose,\xd/black!45/4/Program,\xe/cGrn/5/Audit}{
  \node[num=\c] at (\x-0.75,2.35) {\i};
  \node[ttl, anchor=west] at (\x-0.5,2.35) {\t};}
\draw[flow] (\xa+1.15,0.9) -- (\xb-0.85,0.9);
\draw[flow] (\xb+0.85,0.9) -- (\xc-1.3,0.9);
\draw[flow] (\xc+1.45,0.9) -- (\xd-1.3,0.9);
\draw[flow] (\xd+1.3,0.9) -- (\xe-1.0,0.9);
\draw[flow] (\xe+1.0,0.9) -- (\xe+1.08,0.9) -- (\xe+1.08,2.85) -- (\xa-1.35,2.85) -- (\xa-1.35,0.9) -- (\xa-1.15,0.9);
\node[lab, above] at (\xc,2.85) {archive update: program, exact $(d,A_d)$};
\begin{scope}[shift={(1.4,-2.2)}]
  \foreach \x in {-0.3,0.3}{\fill[cRedF, draw=cRed, line width=0.5pt] (\x,0.42) circle (0.16);
    \fill[cRedF, draw=cRed, line width=0.5pt] (\x-0.27,-0.1) .. controls (\x-0.27,0.27) and (\x+0.27,0.27) .. (\x+0.27,-0.1) -- cycle;}
  \node[ttl, anchor=north, text=cRed!80!black] at (0,-0.18) {Humans};
  \node[body, anchor=north, text=cGrey] at (0,-0.48) {target, warm start, notes;\\ pause, fix, resume};
\end{scope}
\draw[hum] (2.3,-1.85) -- node[lab, left] {files} (2.3,-0.05);
\begin{scope}[shift={(5.6,-2.2)}]
  \foreach \i/\c in {0/cGrey!20,1/cGrey!12,2/white}{\fill[\c, draw=cGrey, line width=0.4pt] (-0.5+\i*0.12,0.05+\i*0.1) rectangle ++(0.75,0.5);
    \foreach \y in {0.15,0.25,0.35}{\draw[cGrey!70, line width=0.3pt] (-0.42+\i*0.12,\y+\i*0.1)--(0.05+\i*0.12,\y+\i*0.1);}}
  \draw[cGrey, line width=0.6pt] (-0.7,0.0) -- (-0.6,-0.18) -- (0.6,-0.18) -- (0.7,0.0);
  \node[ttl, anchor=north, text=cGrey!80!black] at (0,-0.28) {Evidence};
  \node[body, anchor=north, text=cGrey] at (0,-0.58) {spectra, failed routes,\\ certificates};
\end{scope}
\draw[evd] (5.6,-1.45) -- node[lab, right] {evidence} (5.6,-0.05);
\begin{scope}[shift={(9.6,-2.2)}]
  \foreach \i/\h/\c/\t in {0/0.62/cBlue!25/QC,1/0.7/cBlue!40/BCH,2/0.56/cBlue!20/X,3/0.74/cBlue!35/AG,4/0.6/cBlue!25/cyclic,5/0.68/cBlue!45/repair,6/0.58/cBlue!20/SAT,7/0.72/cBlue!35/local}{
    \fill[\c, draw=cBlue, line width=0.35pt] (-0.9+\i*0.22,-0.2) rectangle ++(0.18,\h);
    \node[font=\sffamily\fontsize{3.9}{4}\selectfont, rotate=90, anchor=west, text=cBlue!60!black] at (-0.81+\i*0.22,-0.14) {\t};}
  \draw[cBlue, line width=0.8pt] (-1.0,-0.2) -- (0.95,-0.2);
  \node[ttl, anchor=north, text=cBlue!70!black] at (0,-0.28) {Toolkit};
  \node[body, anchor=north, text=cGrey] at (0,-0.58) {87 hand-written strategies};
\end{scope}
\draw[feed] (9.6,-1.45) -- node[lab, right] {available to discover} (9.6,-0.05);
\draw[evd] (12.2,-0.1) -- (12.2,-1.2) -| (6.1,-1.45);
\node[lab] at (9.2,-1.2) {results};
\node[anchor=north, font=\sffamily\fontsize{6}{7}\selectfont, text=black!70] at (5.6,-3.35)
  {\tikz{\node[draw=cBlue,fill=cSea,rounded corners=1pt,minimum width=0.32cm,minimum height=0.18cm,inner sep=0pt]{};}\, search loop and Toolkit\quad
   \tikz{\node[draw=cAmb,fill=cAmbF,rounded corners=1pt,minimum width=0.32cm,minimum height=0.18cm,inner sep=0pt]{};}\, agents\quad
   \tikz{\node[draw=cGrn,fill=cGrnF,rounded corners=1pt,minimum width=0.32cm,minimum height=0.18cm,inner sep=0pt]{};}\, exact evaluation\quad
   \tikz{\draw[hum] (0,0)--(0.45,0);}\, people};
\end{tikzpicture}
\caption{\textbf{Agent-guided proposals in \sys{}.} A parent program
and its code are selected from the archive (1). The proposal context
holds the code's exact $(d,A_d)$, the island's construction approach,
and the results of earlier attempts (2). Discover uses three
successive model calls to produce a search program; Invent provides an
alternative without direct access to the strategy Toolkit (3). The
generated program searches candidate matrices, using Toolkit routines
when permitted (4). The evaluator checks the returned matrix and
computes the exact weight distribution of valid outputs (5).
Evaluation results update the archive and inform later proposals.
Researchers supply starting points and mathematical observations
through shared files.}
\label{fig:framework}
\end{figure}

\subsection{Agent-Guided Program Generation}
\label{sec:agents}
Each proposal receives the parent program and generator matrix, its
verified weight distribution, the target parameters and distances,
applicable bounds, and the island's construction family. Previous
results, failed attempts, and mathematical observations such as
obstruction bounds and expert notes provide additional context.

\begin{figure}[htbp]
\centering
\begin{tikzpicture}[
 font=\sffamily\small,
 agbox/.style={draw=orange!65!black,fill=orange!7,rounded corners=3pt,
 text width=3.05cm,minimum height=1.15cm,align=left,inner sep=7pt},
 agflow/.style={-{Stealth[length=2mm]},draw=black!55,line width=.8pt},
 aglabel/.style={font=\sffamily\scriptsize,text=black!65,fill=white,inner sep=2pt},
 agshared/.style={draw=black!30,fill=black!3,rounded corners=3pt,align=center,inner sep=6pt}
]
\node[agshared,text width=11.9cm] (context) at (6,2.1)
 {\textbf{Proposal context}\quad Parent program and matrix; verified statistics; targets; evidence};
\node[agbox] (oracle) at (1.8,.3)
 {\textbf{Oracle}\\Diagnose the code\\Suggest a move};
\node[agbox] (strategist) at (6,.3)
 {\textbf{Strategist}\\Compare routes\\Specify a plan};
\node[agbox] (implementer) at (10.2,.3)
 {\textbf{Implementer}\\Write the program\\Use Toolkit};
\node[aglabel,anchor=south west,text=orange!65!black] at (oracle.north west) {DISCOVER};
\draw[agflow] (context.south -| oracle.north) -- (oracle.north);
\draw[agflow] (oracle) -- node[aglabel,above]{diagnosis} (strategist);
\draw[agflow] (strategist) -- node[aglabel,above]{plan} (implementer);
\node[agbox,draw=blue!55!black,fill=blue!4,text width=5.0cm] (invent) at (3,-1.65)
 {\textbf{Invent}\\Describe and implement a procedure\\No Toolkit access};
\node[agshared,text width=4.9cm,minimum height=1.15cm] (checks) at (9.5,-1.65)
 {\textbf{Check and execute}\\Static checks + novelty\\Run with a time limit};
\draw[agflow,draw=blue!55!black] (context.west) -- (-.55,2.1) |- (invent.west);
\draw[agflow] (invent.east) -- (checks.west);
\draw[agflow] (implementer.south) -- (implementer.south |- checks.north);
\node[agshared,draw=green!45!black,fill=green!5,text width=4.9cm] (audit) at (9.5,-3.15)
 {\textbf{Exact evaluation}\\Validate matrix; compute exact $(d,A_d)$};
\draw[agflow] (checks) -- (audit);
\node[align=left,text width=5.0cm,font=\sffamily\scriptsize,text=black!65] at (3,-3.1)
 {Format checks between calls.\\At most one plan-repair attempt.\\Discover and Invent are alternative routes.};
\end{tikzpicture}
\caption{The Discover and Invent proposal operators. Discover
separates proposal generation into diagnosis, planning, and
implementation. Invent develops a procedure without direct access to
Toolkit. Both operators are subject to proposal checks and a
time-limited execution. Valid returned matrices are evaluated exactly;
failure handling is described in the text.}
\label{fig:agent-generation}
\end{figure}
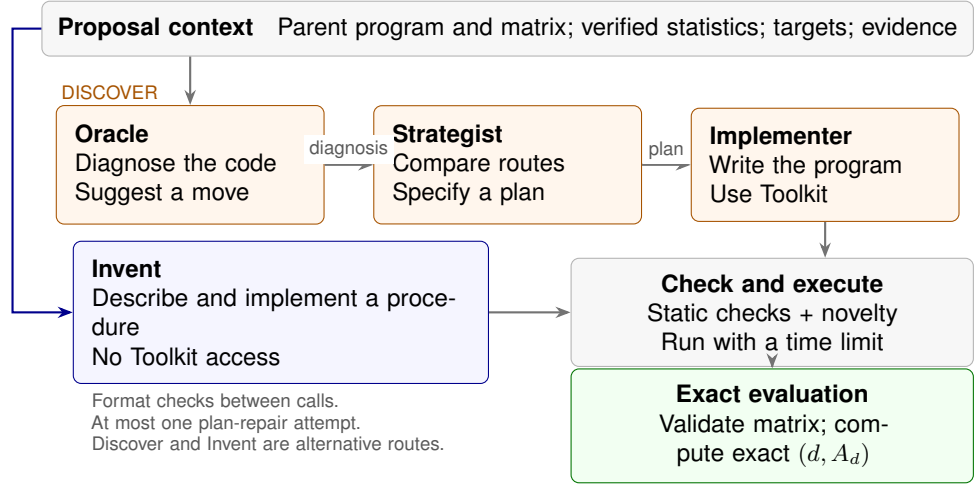

\paragraph{Discover.}
Three successive model calls turn this context into a search program
(Figure~\ref{fig:agent-generation}). The \emph{oracle} diagnoses the
current code and suggests a move, supported by evidence and a condition
for stopping an unsuccessful attempt. The \emph{strategist} compares
construction approaches, checks the parameter arithmetic of proposed
parent codes, and selects one; it records the approach, the move,
whether to modify the parent or start again, the acceptance rule, and
the stopping condition in one structured plan. The \emph{implementer}
translates this plan into a program that can call the strategy Toolkit and
available exact tools. The program must restate the plan and follow its
candidate-selection rule.

\paragraph{Invent.}
A separate operator proposes a procedure without seeing the strategy
catalogue or calling its strategies. It first describes the
representation, the move, the verification step, and the stopping
condition, then implements the procedure from lower-level operations.
It still receives the parent program and accumulated evidence, which
may reflect earlier use of the strategy Toolkit.

\paragraph{Operator selection.}
Outside the invention island, proposals are sampled from the inherited
\emph{diff}, \emph{full}, and \emph{cross} operators, alongside
\emph{discover} and \emph{invent}. Their probabilities adapt to the
parent's distance and recent outcomes, with additional weight given to
\emph{invent} when the parent is below the distance floor set for the
campaign and recent proposals repeatedly return the same code.

\paragraph{Checks and execution.}
Each model call must return its required fields. If the strategist's
plan is incomplete, it gets one more call to complete it, and the
proposal is dropped if it is still incomplete. These checks concern
the format of the plan, not whether it is mathematically sound.
Before a program runs, it is rejected if it only wraps one Toolkit
strategy, if it lacks the required header, or if it re-implements the
exhaustive enumeration; the novelty judge from ShinkaEvolve,
re-prompted for this task, also rejects programs that hardly differ
from the parent. A program that passes runs once on the parent's
matrix, in its own process, under a time limit. Its output counts only
if it is a binary $k\times n$ matrix of rank $k$; the evaluator then
enumerates the code and records $(d,A_d)$. If the program crashes,
times out, or returns anything else, it is scored as if it had
returned the parent unchanged: it keeps the parent's score, does not
enter the archive, and is no longer used as a parent once any program
has reached the distance floor. The archive holds up to 96 programs,
one per distinct code. A new code is added while there is room; when
the archive is full it replaces an archived program under a rule that
keeps small construction families represented
. A code already in the archive is not
added again.

\subsection{Fitness}
\label{sec:fitness}
The fitness of a program is the pair $(d,-A_d)$ of the code it returns,
maximized in lexicographic order: a larger minimum distance $d$ is preferred,
and at equal distance, fewer minimum-weight codewords $A_d$ win. Both fitness
values are exact, from an enumeration of the weight distribution using rbouwer zimmeand alh.

We use a strictly smaller $A_d$ not only as a heuristic to guide further search
but also because it yields a provably more reliable code. Specifically, the
undetected-error probability on the binary symmetric channel with crossover
probability $p$ is
\[
  P_{\mathrm{err}}(C;p) \;=\; \sum_{i=1}^{n} A_i\, p^{i}(1-p)^{n-i},
\]
which represents the probability that the channel turns a transmitted codeword
into another codeword. For small values of $p$, this probability is heavily
dominated by the leading term $A_d p^d(1-p)^{n-d}$, meaning minimizing $A_d$
directly minimizes the likelihood of undetected errors.

\subsection{Initialization: Generation 0 }
\label{sec:gen0}

Our search is initialized with a strong initial program that outputs the linear code with the minimum distance exceeding the Gilbert-Varshamov (GV) bound. To construct this base, we evaluated a diverse portfolio of structured code families, including design-based, quasi-cyclic, and algebraic constructions. Candidate codes are first ranked using computationally inexpensive distance estimates [citation needed], and the most promising candidates undergo an exact evaluation using the Brouwer-Zimmermann algorithm implemented in SageMath. We select the code with the largest minimum distance, using the number of low-weight codewords as a secondary filter.

Table~\ref{tab:seed} compares the selected seeds against the GV bound and random full-rank matrices, reporting both the best and the median minimum distances across 50 random draws. For instances with $k \le 30$, exact distances are rigorously verified via full enumeration. This initialization establishes a highly competitive baseline from which the subsequent search phase seeks further improvements.

\subsection{Specialised Islands}
\label{sec:islands}
We use ShinkaEvolve's island model to explore several approaches to
code construction in parallel. The search begins with eight
populations of programs. Each island receives a short prompt
describing a construction approach, such as algebraic codes over
extension fields, sparse-graph parity-check design, or quasi-cyclic
polynomial matrices. The agents use this perspective when proposing
changes to the parent code; strategies of other families remain
available to them as fallbacks and for final repair.

Two islands have a different role. The cold-start island constructs
codes directly from the target parameters $(n,k)$, using the current
best matrix only as a fallback. The invention island always uses the
\emph{invent} operator, which has no direct access to the 
Toolkit. These islands allow the search to explore alternatives to both
the current construction and the available Toolkit routines.

Programs migrate between islands every six generations, while the
best program on each island is retained. If the best score has not
improved for six generations, a new island is seeded from an archived
program. This arrangement allows promising programs to spread while
maintaining separate searches guided by different construction
approaches. 

\subsection{The Toolkit}
\label{sec:Toolkit}
The is where most of the human knowledge enters. It is a library
of 87 hand-written functions, each taking the current generator matrix
and returning a candidate, so a generated program can invoke any of
them in one line and combine them freely. They range from general
search heuristics, such as simulated annealing and genetic population
search, to algebraic constructions from the coding-theory literature:
cyclic and BCH parent codes, Construction X, quasi-cyclic and
quasi-twisted matrices, and procedures that repair the minimum-weight
codewords of an existing code. Six family cores group these strategies
by the kind of construction they belong to; a core selects suitable
strategies for its input, runs them under a deadline, and returns the
best finalist.

Generated programs can combine these strategies in a search loop,
choose when to apply them, and decide which candidates to keep. They
can also implement additional operations directly. Programs that call
only a single strategy without adding search logic are rejected before
execution. 

\begin{table}[h]
\centering\footnotesize
\setlength{\tabcolsep}{5pt}
\begin{tabular}{lcccccl r}
\toprule
cell & LB & GV & random best & random median $d$ & gen 0 & lane & s \\
\midrule
$\cnk{41}{12}$ & 14 & 11 & $(11,5)$ & 10 & $\mathbf{(12,11)}$ & construction\_core & 214 \\
$\cnk{48}{16}$ & 16 & 11 & $(11,3)$ & 10 & $\mathbf{(14,256)}$ & qc\_core & 321 \\
$\cnk{50}{20}$ & 13 & 10 & $(10,6)$ & 9 & $\mathbf{(11,30)}$ & qc\_core & 413 \\
$\cnk{60}{20}$ & 17 & 13 & $(13,1)$ & 12 & $\mathbf{(16,245)}$ & qc\_core & 307\\
\bottomrule
\end{tabular}
\caption{Generation-0 against random full-rank matrices, exact
$(d,A_d)$. Random: best of 50
draws, and the median $d$ of the 50. GV: the Gilbert--Varshamov
guarantee for $(n,k)$. the portfolio's pick and its winning lane.
}
\label{tab:seed}
\end{table}

\section{Results}
\subsection{Choosing Target Parameters $(n,k)$: Human Intuition}
The framework conducts the search at fixed parameters $(n,k)$. We chose these
parameters outside the evolutionary loop by examining the code tables,
the constructions behind their lower bounds, and the cost of computing the minimum distance.
We considered open entries, where there is a gap between the lower and the upper bound
.
A large gap leaves room for improvement, but may reflect a loose upper
bound rather than a weak construction. We therefore used it only as an
initial filter.

\paragraph{Plateaus and inherited bounds.}
We looked for lengths over which the recorded distance stays constant
as $n$ increases. At $k=21$, the lower bound is $64$ throughout
$165\le n\le173$ and $68$ throughout $179\le n\le183$
 Such plateaus in the minimum distance can arise when a known code is
padded with zero coordinates, leaving open the possibility of using
the additional coordinates more effectively.
\paragraph{Our number theoretic hint.}
Our human input was a hint about where to point the search. The row $k=21$
is frozen at $d=64$ for $165\le n\le173$, around $n=170=2\cdot85$. For block
length $85$, $x^{85}-1$ splits into small factors ($x+1$, $\Phi_5$ and ten
octics), so $k=21=1+4+8+8$ is a natural dimension for an index-2
quasi-cyclic code of length $170$. We therefore suggested a quasi-cyclic core
of length $170$ as a starting point. Similarly, $189=3\cdot63$ with
$22=1+3+6+6+6$ suggested a starting point for $[189,22]$.

\subsection{New Records}
\label{New Records}
\sys{} finds seven binary linear codes that improve the lower bounds recorded
in the CodeTables snapshot of 25~September~2026. Table~\ref{tab:new-records}
summarizes the constructions and their gains. The first five arise from
index-2 quasi-cyclic parent codes of length $170$ and block length $85$;
$[189,22,72]$ is an index-3 quasi-cyclic code with block length $63$.
The $[200,21,77]$ code is obtained by puncturing a listed
$[240,21,104]$ binary Goppa-code construction.

\begin{table}[h]
\centering
\caption{Seven constructions improving CodeTables lower bounds in the
25~September~2026 snapshot. The gain is the difference between the
new minimum distance and the previous lower bound.}
\label{tab:new-records}
\begin{tabular}{lrr}
\toprule
Code $[n,k,d]$ & Previous bound & Gain \\
\midrule
$[172,21,66]$ & 64 & $+2$ \\
$[173,20,68]$ & 67 & $+1$ \\
$[176,21,68]$ & 66 & $+2$ \\
$[181,21,70]$ & 68 & $+2$ \\
$[184,21,72]$ & 69 & $+3$ \\
$[189,22,72]$ & 71 & $+1$ \\
$[200,21,77]$ & 76 & $+1$ \\
\bottomrule
\end{tabular}
\end{table}

The constructions have effects beyond their own parameter pairs.
Explicit derivations yield improvements in $22$ CodeTables entries overall,
with gains of up to three in minimum distance. 

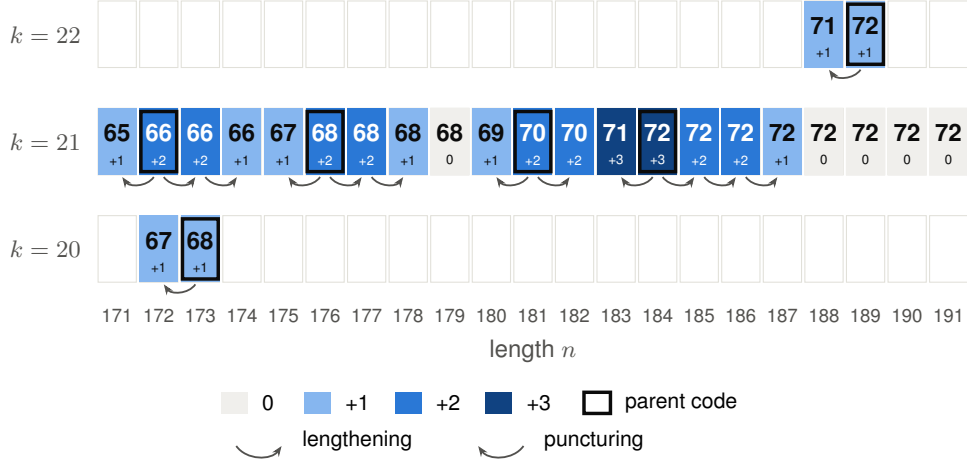
\begin{figure}[h]
\centering

\begin{tikzpicture}[x=0.55cm, y=1.42cm, font=\sffamily]
\node[anchor=east, text=inkS, font=\sffamily\small] at (170.35,2.00) {$k=22$};
\draw[gridc, line width=0.4pt] (170.54,1.69) rectangle (171.46,2.31);
\draw[gridc, line width=0.4pt] (171.54,1.69) rectangle (172.46,2.31);
\draw[gridc, line width=0.4pt] (172.54,1.69) rectangle (173.46,2.31);
\draw[gridc, line width=0.4pt] (173.54,1.69) rectangle (174.46,2.31);
\draw[gridc, line width=0.4pt] (174.54,1.69) rectangle (175.46,2.31);
\draw[gridc, line width=0.4pt] (175.54,1.69) rectangle (176.46,2.31);
\draw[gridc, line width=0.4pt] (176.54,1.69) rectangle (177.46,2.31);
\draw[gridc, line width=0.4pt] (177.54,1.69) rectangle (178.46,2.31);
\draw[gridc, line width=0.4pt] (178.54,1.69) rectangle (179.46,2.31);
\draw[gridc, line width=0.4pt] (179.54,1.69) rectangle (180.46,2.31);
\draw[gridc, line width=0.4pt] (180.54,1.69) rectangle (181.46,2.31);
\draw[gridc, line width=0.4pt] (181.54,1.69) rectangle (182.46,2.31);
\draw[gridc, line width=0.4pt] (182.54,1.69) rectangle (183.46,2.31);
\draw[gridc, line width=0.4pt] (183.54,1.69) rectangle (184.46,2.31);
\draw[gridc, line width=0.4pt] (184.54,1.69) rectangle (185.46,2.31);
\draw[gridc, line width=0.4pt] (185.54,1.69) rectangle (186.46,2.31);
\draw[gridc, line width=0.4pt] (186.54,1.69) rectangle (187.46,2.31);
\draw[gridc, line width=0.4pt] (187.54,1.69) rectangle (188.46,2.31);
\draw[gridc, line width=0.4pt] (188.54,1.69) rectangle (189.46,2.31);
\draw[gridc, line width=0.4pt] (189.54,1.69) rectangle (190.46,2.31);
\draw[gridc, line width=0.4pt] (190.54,1.69) rectangle (191.46,2.31);
\node[anchor=east, text=inkS, font=\sffamily\small] at (170.35,1.00) {$k=21$};
\draw[gridc, line width=0.4pt] (170.54,0.69) rectangle (171.46,1.31);
\draw[gridc, line width=0.4pt] (171.54,0.69) rectangle (172.46,1.31);
\draw[gridc, line width=0.4pt] (172.54,0.69) rectangle (173.46,1.31);
\draw[gridc, line width=0.4pt] (173.54,0.69) rectangle (174.46,1.31);
\draw[gridc, line width=0.4pt] (174.54,0.69) rectangle (175.46,1.31);
\draw[gridc, line width=0.4pt] (175.54,0.69) rectangle (176.46,1.31);
\draw[gridc, line width=0.4pt] (176.54,0.69) rectangle (177.46,1.31);
\draw[gridc, line width=0.4pt] (177.54,0.69) rectangle (178.46,1.31);
\draw[gridc, line width=0.4pt] (178.54,0.69) rectangle (179.46,1.31);
\draw[gridc, line width=0.4pt] (179.54,0.69) rectangle (180.46,1.31);
\draw[gridc, line width=0.4pt] (180.54,0.69) rectangle (181.46,1.31);
\draw[gridc, line width=0.4pt] (181.54,0.69) rectangle (182.46,1.31);
\draw[gridc, line width=0.4pt] (182.54,0.69) rectangle (183.46,1.31);
\draw[gridc, line width=0.4pt] (183.54,0.69) rectangle (184.46,1.31);
\draw[gridc, line width=0.4pt] (184.54,0.69) rectangle (185.46,1.31);
\draw[gridc, line width=0.4pt] (185.54,0.69) rectangle (186.46,1.31);
\draw[gridc, line width=0.4pt] (186.54,0.69) rectangle (187.46,1.31);
\draw[gridc, line width=0.4pt] (187.54,0.69) rectangle (188.46,1.31);
\draw[gridc, line width=0.4pt] (188.54,0.69) rectangle (189.46,1.31);
\draw[gridc, line width=0.4pt] (189.54,0.69) rectangle (190.46,1.31);
\draw[gridc, line width=0.4pt] (190.54,0.69) rectangle (191.46,1.31);
\node[anchor=east, text=inkS, font=\sffamily\small] at (170.35,0.00) {$k=20$};
\draw[gridc, line width=0.4pt] (170.54,-0.31) rectangle (171.46,0.31);
\draw[gridc, line width=0.4pt] (171.54,-0.31) rectangle (172.46,0.31);
\draw[gridc, line width=0.4pt] (172.54,-0.31) rectangle (173.46,0.31);
\draw[gridc, line width=0.4pt] (173.54,-0.31) rectangle (174.46,0.31);
\draw[gridc, line width=0.4pt] (174.54,-0.31) rectangle (175.46,0.31);
\draw[gridc, line width=0.4pt] (175.54,-0.31) rectangle (176.46,0.31);
\draw[gridc, line width=0.4pt] (176.54,-0.31) rectangle (177.46,0.31);
\draw[gridc, line width=0.4pt] (177.54,-0.31) rectangle (178.46,0.31);
\draw[gridc, line width=0.4pt] (178.54,-0.31) rectangle (179.46,0.31);
\draw[gridc, line width=0.4pt] (179.54,-0.31) rectangle (180.46,0.31);
\draw[gridc, line width=0.4pt] (180.54,-0.31) rectangle (181.46,0.31);
\draw[gridc, line width=0.4pt] (181.54,-0.31) rectangle (182.46,0.31);
\draw[gridc, line width=0.4pt] (182.54,-0.31) rectangle (183.46,0.31);
\draw[gridc, line width=0.4pt] (183.54,-0.31) rectangle (184.46,0.31);
\draw[gridc, line width=0.4pt] (184.54,-0.31) rectangle (185.46,0.31);
\draw[gridc, line width=0.4pt] (185.54,-0.31) rectangle (186.46,0.31);
\draw[gridc, line width=0.4pt] (186.54,-0.31) rectangle (187.46,0.31);
\draw[gridc, line width=0.4pt] (187.54,-0.31) rectangle (188.46,0.31);
\draw[gridc, line width=0.4pt] (188.54,-0.31) rectangle (189.46,0.31);
\draw[gridc, line width=0.4pt] (189.54,-0.31) rectangle (190.46,0.31);
\draw[gridc, line width=0.4pt] (190.54,-0.31) rectangle (191.46,0.31);
\node[text=inkS, font=\sffamily\scriptsize] at (171,-0.62) {171};
\node[text=inkS, font=\sffamily\scriptsize] at (172,-0.62) {172};
\node[text=inkS, font=\sffamily\scriptsize] at (173,-0.62) {173};
\node[text=inkS, font=\sffamily\scriptsize] at (174,-0.62) {174};
\node[text=inkS, font=\sffamily\scriptsize] at (175,-0.62) {175};
\node[text=inkS, font=\sffamily\scriptsize] at (176,-0.62) {176};
\node[text=inkS, font=\sffamily\scriptsize] at (177,-0.62) {177};
\node[text=inkS, font=\sffamily\scriptsize] at (178,-0.62) {178};
\node[text=inkS, font=\sffamily\scriptsize] at (179,-0.62) {179};
\node[text=inkS, font=\sffamily\scriptsize] at (180,-0.62) {180};
\node[text=inkS, font=\sffamily\scriptsize] at (181,-0.62) {181};
\node[text=inkS, font=\sffamily\scriptsize] at (182,-0.62) {182};
\node[text=inkS, font=\sffamily\scriptsize] at (183,-0.62) {183};
\node[text=inkS, font=\sffamily\scriptsize] at (184,-0.62) {184};
\node[text=inkS, font=\sffamily\scriptsize] at (185,-0.62) {185};
\node[text=inkS, font=\sffamily\scriptsize] at (186,-0.62) {186};
\node[text=inkS, font=\sffamily\scriptsize] at (187,-0.62) {187};
\node[text=inkS, font=\sffamily\scriptsize] at (188,-0.62) {188};
\node[text=inkS, font=\sffamily\scriptsize] at (189,-0.62) {189};
\node[text=inkS, font=\sffamily\scriptsize] at (190,-0.62) {190};
\node[text=inkS, font=\sffamily\scriptsize] at (191,-0.62) {191};
\node[text=inkS, font=\sffamily\small] at (181,-0.95) {length $n$};
\fill[rampA] (170.54,0.69) rectangle (171.46,1.31);
\node[text=inkP, font=\sffamily\small\bfseries] at (171,1.08) {65};
\node[text=inkP, font=\sffamily\tiny] at (171,0.83) {+1};
\fill[rampA] (171.54,-0.31) rectangle (172.46,0.31);
\node[text=inkP, font=\sffamily\small\bfseries] at (172,0.08) {67};
\node[text=inkP, font=\sffamily\tiny] at (172,-0.17) {+1};
\fill[rampB] (171.54,0.69) rectangle (172.46,1.31);
\draw[inkP, line width=1.3pt] (171.57,0.72) rectangle (172.43,1.28);
\node[text=white, font=\sffamily\small\bfseries] at (172,1.08) {66};
\node[text=white, font=\sffamily\tiny] at (172,0.83) {+2};
\fill[rampA] (172.54,-0.31) rectangle (173.46,0.31);
\draw[inkP, line width=1.3pt] (172.57,-0.28) rectangle (173.43,0.28);
\node[text=inkP, font=\sffamily\small\bfseries] at (173,0.08) {68};
\node[text=inkP, font=\sffamily\tiny] at (173,-0.17) {+1};
\fill[rampB] (172.54,0.69) rectangle (173.46,1.31);
\node[text=white, font=\sffamily\small\bfseries] at (173,1.08) {66};
\node[text=white, font=\sffamily\tiny] at (173,0.83) {+2};
\fill[rampA] (173.54,0.69) rectangle (174.46,1.31);
\node[text=inkP, font=\sffamily\small\bfseries] at (174,1.08) {66};
\node[text=inkP, font=\sffamily\tiny] at (174,0.83) {+1};
\fill[rampA] (174.54,0.69) rectangle (175.46,1.31);
\node[text=inkP, font=\sffamily\small\bfseries] at (175,1.08) {67};
\node[text=inkP, font=\sffamily\tiny] at (175,0.83) {+1};
\fill[rampB] (175.54,0.69) rectangle (176.46,1.31);
\draw[inkP, line width=1.3pt] (175.57,0.72) rectangle (176.43,1.28);
\node[text=white, font=\sffamily\small\bfseries] at (176,1.08) {68};
\node[text=white, font=\sffamily\tiny] at (176,0.83) {+2};
\fill[rampB] (176.54,0.69) rectangle (177.46,1.31);
\node[text=white, font=\sffamily\small\bfseries] at (177,1.08) {68};
\node[text=white, font=\sffamily\tiny] at (177,0.83) {+2};
\fill[rampA] (177.54,0.69) rectangle (178.46,1.31);
\node[text=inkP, font=\sffamily\small\bfseries] at (178,1.08) {68};
\node[text=inkP, font=\sffamily\tiny] at (178,0.83) {+1};
\fill[neutral] (178.54,0.69) rectangle (179.46,1.31);
\node[text=inkP, font=\sffamily\small\bfseries] at (179,1.08) {68};
\node[text=inkP, font=\sffamily\tiny] at (179,0.83) {0};
\fill[rampA] (179.54,0.69) rectangle (180.46,1.31);
\node[text=inkP, font=\sffamily\small\bfseries] at (180,1.08) {69};
\node[text=inkP, font=\sffamily\tiny] at (180,0.83) {+1};
\fill[rampB] (180.54,0.69) rectangle (181.46,1.31);
\draw[inkP, line width=1.3pt] (180.57,0.72) rectangle (181.43,1.28);
\node[text=white, font=\sffamily\small\bfseries] at (181,1.08) {70};
\node[text=white, font=\sffamily\tiny] at (181,0.83) {+2};
\fill[rampB] (181.54,0.69) rectangle (182.46,1.31);
\node[text=white, font=\sffamily\small\bfseries] at (182,1.08) {70};
\node[text=white, font=\sffamily\tiny] at (182,0.83) {+2};
\fill[rampC] (182.54,0.69) rectangle (183.46,1.31);
\node[text=white, font=\sffamily\small\bfseries] at (183,1.08) {71};
\node[text=white, font=\sffamily\tiny] at (183,0.83) {+3};
\fill[rampC] (183.54,0.69) rectangle (184.46,1.31);
\draw[inkP, line width=1.3pt] (183.57,0.72) rectangle (184.43,1.28);
\node[text=white, font=\sffamily\small\bfseries] at (184,1.08) {72};
\node[text=white, font=\sffamily\tiny] at (184,0.83) {+3};
\fill[rampB] (184.54,0.69) rectangle (185.46,1.31);
\node[text=white, font=\sffamily\small\bfseries] at (185,1.08) {72};
\node[text=white, font=\sffamily\tiny] at (185,0.83) {+2};
\fill[rampB] (185.54,0.69) rectangle (186.46,1.31);
\node[text=white, font=\sffamily\small\bfseries] at (186,1.08) {72};
\node[text=white, font=\sffamily\tiny] at (186,0.83) {+2};
\fill[rampA] (186.54,0.69) rectangle (187.46,1.31);
\node[text=inkP, font=\sffamily\small\bfseries] at (187,1.08) {72};
\node[text=inkP, font=\sffamily\tiny] at (187,0.83) {+1};
\fill[neutral] (187.54,0.69) rectangle (188.46,1.31);
\node[text=inkP, font=\sffamily\small\bfseries] at (188,1.08) {72};
\node[text=inkP, font=\sffamily\tiny] at (188,0.83) {0};
\fill[rampA] (187.54,1.69) rectangle (188.46,2.31);
\node[text=inkP, font=\sffamily\small\bfseries] at (188,2.08) {71};
\node[text=inkP, font=\sffamily\tiny] at (188,1.83) {+1};
\fill[neutral] (188.54,0.69) rectangle (189.46,1.31);
\node[text=inkP, font=\sffamily\small\bfseries] at (189,1.08) {72};
\node[text=inkP, font=\sffamily\tiny] at (189,0.83) {0};
\fill[rampA] (188.54,1.69) rectangle (189.46,2.31);
\draw[inkP, line width=1.3pt] (188.57,1.72) rectangle (189.43,2.28);
\node[text=inkP, font=\sffamily\small\bfseries] at (189,2.08) {72};
\node[text=inkP, font=\sffamily\tiny] at (189,1.83) {+1};
\fill[neutral] (189.54,0.69) rectangle (190.46,1.31);
\node[text=inkP, font=\sffamily\small\bfseries] at (190,1.08) {72};
\node[text=inkP, font=\sffamily\tiny] at (190,0.83) {0};
\fill[neutral] (190.54,0.69) rectangle (191.46,1.31);
\node[text=inkP, font=\sffamily\small\bfseries] at (191,1.08) {72};
\node[text=inkP, font=\sffamily\tiny] at (191,0.83) {0};
\draw[-{Stealth[length=4pt]}, inkS, line width=0.7pt] (171.88,0.67) to[bend left=55] (171.12,0.67);
\draw[-{Stealth[length=4pt]}, inkS, line width=0.7pt] (172.88,-0.33) to[bend left=55] (172.12,-0.33);
\draw[-{Stealth[length=4pt]}, inkS, line width=0.7pt] (172.12,0.67) to[bend right=55] (172.88,0.67);
\draw[-{Stealth[length=4pt]}, inkS, line width=0.7pt] (173.12,0.67) to[bend right=55] (173.88,0.67);
\draw[-{Stealth[length=4pt]}, inkS, line width=0.7pt] (175.88,0.67) to[bend left=55] (175.12,0.67);
\draw[-{Stealth[length=4pt]}, inkS, line width=0.7pt] (176.12,0.67) to[bend right=55] (176.88,0.67);
\draw[-{Stealth[length=4pt]}, inkS, line width=0.7pt] (177.12,0.67) to[bend right=55] (177.88,0.67);
\draw[-{Stealth[length=4pt]}, inkS, line width=0.7pt] (180.88,0.67) to[bend left=55] (180.12,0.67);
\draw[-{Stealth[length=4pt]}, inkS, line width=0.7pt] (181.12,0.67) to[bend right=55] (181.88,0.67);
\draw[-{Stealth[length=4pt]}, inkS, line width=0.7pt] (183.88,0.67) to[bend left=55] (183.12,0.67);
\draw[-{Stealth[length=4pt]}, inkS, line width=0.7pt] (184.12,0.67) to[bend right=55] (184.88,0.67);
\draw[-{Stealth[length=4pt]}, inkS, line width=0.7pt] (185.12,0.67) to[bend right=55] (185.88,0.67);
\draw[-{Stealth[length=4pt]}, inkS, line width=0.7pt] (186.12,0.67) to[bend right=55] (186.88,0.67);
\draw[-{Stealth[length=4pt]}, inkS, line width=0.7pt] (188.88,1.67) to[bend left=55] (188.12,1.67);
\end{tikzpicture}
\\[4pt]
\begin{tikzpicture}[font=\sffamily\footnotesize]
\fill[neutral] (0,0) rectangle (0.35,0.28);
\node[anchor=west] at (0.42,0.14) {0};

\fill[rampA] (1.1,0) rectangle (1.45,0.28);
\node[anchor=west] at (1.52,0.14) {+1};

\fill[rampB] (2.3,0) rectangle (2.65,0.28);
\node[anchor=west] at (2.72,0.14) {+2};

\fill[rampC] (3.5,0) rectangle (3.85,0.28);
\node[anchor=west] at (3.92,0.14) {+3};

\draw[inkP, line width=1.3pt] (4.8,0) rectangle (5.15,0.28);
\node[anchor=west] at (5.22,0.14) {parent code};

\draw[-{Stealth[length=4pt]}, inkS, line width=0.7pt]
  (0.2,-0.43) to[bend right=55] (0.8,-0.43);
\node[anchor=west] at (0.95,-0.36) {lengthening};

\draw[-{Stealth[length=4pt]}, inkS, line width=0.7pt]
  (4.0,-0.43) to[bend left=55] (3.4,-0.43);
\node[anchor=west] at (4.15,-0.36) {puncturing};
\end{tikzpicture}
\caption{Codes at lengths $171$-$191$. Cells show minimum distance
and gain over CodeTables; outlines identify parent codes.}
\label{fig:map}
\end{figure}

Figure~\ref{fig:map} shows
the cluster at lengths $171$-$191$; the $[200,21,77]$ construction lies
outside its displayed range. 

We verified the reported minimum distances using SageMath
\citep{sagemath} and Magma \citep{bosma1997magma}. The appendix and
accompanying files provide the generator matrices, weight distributions,
derivations, and verification scripts.

Markus Grassl, who maintains \href{https://www.codetables.de}{CodeTables} ,
independently checked the new constructions and accepted them for
inclusion. The live tables have not yet been updated. The improvements are measured against the bounds listed in CodeTables
on 24~September~2026.
\subsection{Structural Advantages of Our Codes}
\paragraph{A simpler construction matching the best-known lower bound.}
\sys{} also finds codes that match existing distance bounds while admitting concise algebraic descriptions. For example, it constructs a binary $[50,25,10]$ code, meeting the current best-known lower bound of $10$ (the upper bound is $12$). The code is pure double circulant: its generator matrix has the form $[\,I_{25}\mid A\,]$, where $A$ is a $25\times25$ circulant matrix. This example shows that the search can recover structured, readily reproducible constructions even when it does not improve the distance bound.

\paragraph{Comparison with direct prompting of GPT-6 Astra.}

For $(n,k)=(172,21)$, \sys{} finds a code with minimum distance $66$
built from a quasi-cyclic $[170,21,64]$ core by adding two coordinates
via Construction~X. The best code obtained by directly prompting
GPT-6 Astra at extra-high reasoning effort in our comparison has
minimum distance $65$.

At $(n,k)=(200,21)$, both approaches reach distance $77$, but they produce inequivalent codes with different weight distributions. The \sys{} code has $25$ minimum-weight codewords, compared with $75$ for the directly prompted code. It also has fewer nonzero codewords of weight at most $78$ ($264$ versus $303$). 

Consequently, its parity extension yields a $[201,21,78]$ code with $264$ minimum-weight codewords, compared with $303$ for the extension of the other code.

Direct prompting might recover these constructions with more attempts. The contribution of \sys{} is a systematic search process that can be applied across pairs of parameters without creating a new prompt for each target. It preserves verified candidates, uses exact evaluations to guide subsequent proposals, and produces codes whose properties can be independently checked.

\section{Conclusion}
Using \sys{}, we found seven parent codes using this method. The code modification techniques extend these results to 22 parameter pairs. Although we applied \sys{} only to binary codes, the same approach can be used for $q$-ary codes, given an exact minimum-distance evaluator over $\mathbb{F}_q$. Since exact evaluation then scales like $q^k$ rather than $2^k$, the feasible dimensions are correspondingly smaller. The search can also be restricted to structured families to look for the best codes within them.

The time complexity of our method scales exponentially in terms of code dimension. However, for fixed $k$, longer codes remain inexpensive to evaluate, so \sys{} may extend to larger lengths. A practical obstacle is comparison: the binary tables of \href{https://www.codetables.de}{CodeTables} currently stop at length~256, which makes it harder to tell whether a longer code beats the best known code of the same length and dimension.

\section{Resource usage}
We used a single 80GB H100 on the server of the Mathematical Data Science laboratory of EPFL. Each experiment took 3-4 hours on average using these resources.
\section{Acknowledgement}
We thank Gabriel Jimenez Alejandro Calles for their contributions to adapting the ShinkaEvolve database for \sys{}, and Anja Surina for stimulating discussions on evolutionary search.
\section{LLM Usage Statement}
We used LLM-based chatbots to improve and polish the writing of this paper. Given the nature of our research, we also used LLMs for ideation within the framework, for example, to generate improved prompts based on successful prior results. However, beyond the framework itself, we did not use LLMs to generate ideas or strategies for applying the framework.
We also used LLMs to identify potentially relevant citations. All such citations were independently verified by the authors before being included in the literature review. Finally, LLMs were used to assist in drafting portions of the manuscript; these sections were subsequently reviewed, revised, and, where necessary, rewritten by the authors.

\label{refs:start}
\bibliography{references}

@manual{sagemath,
  Key          = {SageMath},
  Author       = {{The Sage Developers}},
  Title        = {{S}ageMath, the {S}age {M}athematics {S}oftware {S}ystem},
  note         = {\url{https://www.sagemath.org}},
  Year         = {2024} 
}

@article{novikov2025alphaevolve,
  title   = {{AlphaEvolve}: A coding agent for scientific and algorithmic discovery},
  author  = {Novikov, Alexander and others},
  journal = {arXiv preprint},
  year    = {2025},
  eprint  = {2506.13131},
  archivePrefix = {arXiv},
  note    = {arXiv:2506.13131}
}

@article{lange2025shinka,
  title   = {{ShinkaEvolve}: Towards open-ended and sample-efficient program evolution},
  author  = {Lange, Robert Tjarko and Imajuku, Yuki and Cetin, Edoardo},
  journal = {arXiv preprint},
  year    = {2025},
  eprint  = {2509.19349},
  archivePrefix = {arXiv},
  note    = {arXiv:2509.19349}
}

@article{lehman2022elm,
  title   = {Evolution through large models},
  author  = {Lehman, Joel and Gordon, Jonathan and Jain, Shawn and Ndousse, Kamal and Yeh, Cathy and Stanley, Kenneth O.},
  journal = {arXiv preprint},
  year    = {2022},
  eprint  = {2206.08896},
  archivePrefix = {arXiv},
  note    = {arXiv:2206.08896}
}

@article{meyerson2023lmx,
  title   = {Language model crossover: Variation through few-shot prompting},
  author  = {Meyerson, Elliot and Nelson, Mark J. and Bradley, Herbie and Gaier, Adam and Moradi, Arash and Hoover, Amy K. and Lehman, Joel},
  journal = {arXiv preprint},
  year    = {2023},
  eprint  = {2302.12170},
  archivePrefix = {arXiv},
  note    = {arXiv:2302.12170}
}

@article{hu2024automated,
  title   = {Automated design of agentic systems},
  author  = {Hu, Shengran and Lu, Cong and Clune, Jeff},
  journal = {arXiv preprint},
  year    = {2024},
  eprint  = {2408.08435},
  archivePrefix = {arXiv},
  note    = {arXiv:2408.08435}
}

@article{lu2024aiscientist,
  title   = {The {AI} {Scientist}: Towards fully automated open-ended scientific discovery},
  author  = {Lu, Chris and Lu, Cong and Lange, Robert Tjarko and Foerster, Jakob and Clune, Jeff and Ha, David},
  journal = {arXiv preprint},
  year    = {2024},
  eprint  = {2408.06292},
  archivePrefix = {arXiv},
  note    = {arXiv:2408.06292}
}

@article{mouret2015illuminating,
  title   = {Illuminating search spaces by mapping elites},
  author  = {Mouret, Jean-Baptiste and Clune, Jeff},
  journal = {arXiv preprint},
  year    = {2015},
  eprint  = {1504.04909},
  archivePrefix = {arXiv},
  note    = {arXiv:1504.04909}
}

@article{alba2002parallelism,
  title   = {Parallelism and evolutionary algorithms},
  author  = {Alba, Enrique and Tomassini, Marco},
  journal = {IEEE Transactions on Evolutionary Computation},
  volume  = {6},
  number  = {5},
  pages   = {443--462},
  year    = {2002},
  doi     = {10.1109/TEVC.2002.800880}
}

@article{auer2002finite,
  title   = {Finite-time analysis of the multiarmed bandit problem},
  author  = {Auer, Peter and Cesa-Bianchi, Nicol{\`o} and Fischer, Paul},
  journal = {Machine Learning},
  volume  = {47},
  number  = {2--3},
  pages   = {235--256},
  year    = {2002},
  doi     = {10.1023/A:1013689704352}
}

@article{wagner2021constructions,
  title   = {Constructions in combinatorics via neural networks},
  author  = {Wagner, Adam Zsolt},
  journal = {arXiv preprint},
  year    = {2021},
  eprint  = {2104.14516},
  archivePrefix = {arXiv},
  note    = {arXiv:2104.14516}
}

@article{charton2024patternboost,
  title   = {{PatternBoost}: Constructions in mathematics with a little help from {AI}},
  author  = {Charton, Fran{\c{c}}ois and Ellenberg, Jordan S. and Wagner, Adam Zsolt and Williamson, Geordie},
  journal = {arXiv preprint},
  year    = {2024},
  eprint  = {2411.00566},
  archivePrefix = {arXiv},
  note    = {arXiv:2411.00566}
}

@article{trinh2024solving,
  title   = {Solving olympiad geometry without human demonstrations},
  author  = {Trinh, Trieu H. and Wu, Yuhuai and Le, Quoc V. and He, He and Luong, Thang},
  journal = {Nature},
  volume  = {625},
  pages   = {476--482},
  year    = {2024},
  doi     = {10.1038/s41586-023-06747-5}
}

@misc{georgiev2025,
      title={Mathematical exploration and discovery at scale}, 
      author={Bogdan Georgiev and Javier Gómez-Serrano and Terence Tao and Adam Zsolt Wagner},
      year={2025},
      eprint={2511.02864},
      archivePrefix={arXiv},
      primaryClass={cs.NE}
}

@article{vardy1997intractability,
  title   = {The intractability of computing the minimum distance of a code},
  author  = {Vardy, Alexander},
  journal = {IEEE Transactions on Information Theory},
  volume  = {43},
  number  = {6},
  pages   = {1757--1766},
  year    = {1997},
  doi     = {10.1109/18.641542}
}

@incollection{grassl2006searching,
  title     = {Searching for linear codes with large minimum distance},
  author    = {Grassl, Markus},
  booktitle = {Discovering Mathematics with Magma},
  editor    = {Bosma, Wieb and Cannon, John},
  publisher = {Springer},
  pages     = {287--313},
  year      = {2006},
  doi       = {10.1007/978-3-540-37634-7_13}
}

@book{huffman2003fundamentals,
  title     = {Fundamentals of Error-Correcting Codes},
  author    = {Huffman, W. Cary and Pless, Vera},
  publisher = {Cambridge University Press},
  year      = {2003},
  doi       = {10.1017/CBO9780511807077}
}

@article{griesmer1960bound,
  title   = {A bound for error-correcting codes},
  author  = {Griesmer, James H.},
  journal = {IBM Journal of Research and Development},
  volume  = {4},
  number  = {5},
  pages   = {532--542},
  year    = {1960},
  doi     = {10.1147/rd.45.0532}
}

@article{prange1962information,
  title   = {The use of information sets in decoding cyclic codes},
  author  = {Prange, Eugene},
  journal = {IRE Transactions on Information Theory},
  volume  = {8},
  number  = {5},
  pages   = {5--9},
  year    = {1962},
  doi     = {10.1109/TIT.1962.1057777}
}

@article{sloane1972new,
  title   = {New binary codes},
  author  = {Sloane, Neil J. A. and Reddy, Sudhakar M. and Chen, Chin-Long},
  journal = {IEEE Transactions on Information Theory},
  volume  = {18},
  number  = {4},
  pages   = {503--510},
  year    = {1972},
  doi     = {10.1109/TIT.1972.1054833}
}

@article{chen1969quasi,
  title   = {Some results on quasi-cyclic codes},
  author  = {Chen, Chin-Long and Peterson, W. Wesley and Weldon, Edward J.},
  journal = {Information and Control},
  volume  = {15},
  number  = {5},
  pages   = {407--423},
  year    = {1969},
  doi     = {10.1016/S0019-9958(69)90497-5}
}

@misc{evotune,
      title={Algorithm Discovery With LLMs: Evolutionary Search Meets Reinforcement Learning}, 
      author={Anja Surina and Amin Mansouri and Lars Quaedvlieg and Amal Seddas and Maryna Viazovska and Emmanuel Abbe and Caglar Gulcehre},
      year={2025},
      eprint={2504.05108},
      archivePrefix={arXiv},
      primaryClass={cs.AI}
}

@article{gulliver1991some,
  title   = {Some best rate $1/p$ and rate $(p-1)/p$ systematic quasi-cyclic codes},
  author  = {Gulliver, T. Aaron and Bhargava, Vijay K.},
  journal = {IEEE Transactions on Information Theory},
  volume  = {37},
  number  = {3},
  pages   = {552--555},
  year    = {1991},
  doi     = {10.1109/18.79911}
}

@article{daskalov2003new,
  title   = {New binary one-generator quasi-cyclic codes},
  author  = {Daskalov, Rumen and Hristov, Plamen},
  journal = {IEEE Transactions on Information Theory},
  volume  = {49},
  number  = {11},
  pages   = {3001--3005},
  year    = {2003},
  doi     = {10.1109/TIT.2003.819337}
}

@book{betten2006error,
  title     = {Error-Correcting Linear Codes: Classification by Isometry and Applications},
  author    = {Betten, Anton and Braun, Michael and Fripertinger, Harald and Kerber, Adalbert and Kohnert, Axel and Wassermann, Alfred},
  publisher = {Springer},
  series    = {Algorithms and Computation in Mathematics},
  volume    = {18},
  year      = {2006},
  doi       = {10.1007/3-540-31703-1}
}

@article{bosma1997magma,
  title   = {The {Magma} algebra system {I}: The user language},
  author  = {Bosma, Wieb and Cannon, John and Playoust, Catherine},
  journal = {Journal of Symbolic Computation},
  volume  = {24},
  number  = {3--4},
  pages   = {235--265},
  year    = {1997},
  doi     = {10.1006/jsco.1996.0125}
}

@article{chen2015iterative,
  author  = {Chen, Eric Zhi},
  title   = {A new iterative computer search algorithm for good quasi-twisted codes},
  journal = {Designs, Codes and Cryptography},
  volume  = {76},
  number  = {2},
  pages   = {307--323},
  year    = {2015},
  doi     = {10.1007/s10623-014-9938-4}
}

@article{chen2015augmentation,
  title   = {New binary $h$-generator quasi-cyclic codes by augmentation and new minimum distance bounds},
  author  = {Chen, Eric Zhi},
  journal = {Designs, Codes and Cryptography},
  year    = {2015},
  doi     = {10.1007/s10623-015-0059-5}
}

@article{aydin2017augmentation,
  title   = {New binary linear codes from quasi-cyclic codes and an augmentation algorithm},
  author  = {Aydin, Nuh and Connolly, Nicholas and Murphree, John},
  journal = {Applicable Algebra in Engineering, Communication and Computing},
  volume  = {28},
  pages   = {339--350},
  year    = {2017},
  doi     = {10.1007/s00200-017-0327-x}
}

@article{aydin2001structure,
  title   = {The structure of 1-generator quasi-twisted codes and new linear codes},
  author  = {Aydin, Nuh and Siap, Irfan and Ray-Chaudhuri, Dijen K.},
  journal = {Designs, Codes and Cryptography},
  volume  = {24},
  number  = {3},
  pages   = {313--326},
  year    = {2001},
  doi     = {10.1023/A:1011283523000}
}

@article{yu2024group,
  title   = {New record-breaking binary linear codes constructed from group codes},
  author  = {Yu, Cong and Zhu, Shixin and Chen, Hao and Li, Yang and Zhang, Xiuyu},
  journal = {arXiv preprint},
  year    = {2024},
  eprint  = {2412.15551},
  archivePrefix = {arXiv},
  note    = {arXiv:2412.15551}
}

@article{wu2025lcd,
  title   = {Linear complementary dual codes constructed from reinforcement learning},
  author  = {Wu, Yansheng and Ma, Jin and Yang, Shangdong},
  journal = {Journal of Systems Science and Complexity},
  volume  = {38},
  pages   = {1388--1403},
  year    = {2025},
  doi     = {10.1007/s11424-025-4313-2}
}

@article{gurkan2026mutation,
  title   = {Mutation without variation: Convergence dynamics in {LLM}-driven program evolution},
  author  = {Gurkan, Can and Stonedahl, Forrest and Wilensky, Uri},
  journal = {arXiv preprint},
  year    = {2026},
  eprint  = {2606.05408},
  archivePrefix = {arXiv},
  note    = {arXiv:2606.05408}
}

@article{wu2024evolutionary,
  title   = {Evolutionary computation in the era of large language model: Survey and roadmap},
  author  = {Wu, Xingyu and Wu, Sheng-hao and Wu, Jibin and Feng, Liang and Tan, Kay Chen},
  journal = {arXiv preprint},
  year    = {2024},
  eprint  = {2401.10034},
  archivePrefix = {arXiv},
  note    = {arXiv:2401.10034}
}

@article{sason2006performance,
  author  = {Sason, Igal and Shamai, Shlomo},
  title   = {Performance Analysis of Linear Codes under
             Maximum-Likelihood Decoding: A Tutorial},
  journal = {Foundations and Trends in Communications
             and Information Theory},
  volume  = {3},
  number  = {1--2},
  pages   = {1--225},
  year    = {2006},
  doi     = {10.1561/0100000009}
}

@article{varshamov1957estimate,
  author    = {Varshamov, Rom R.},
  title     = {Estimate of the number of signals in error correcting codes},
  journal   = {Doklady Akademii Nauk SSSR},
  volume    = {117},
  pages     = {739--741},
  year      = {1957}
}

@article{weindel2025llm,
  title   = {{LLM}-Guided Search for Deletion-Correcting Codes},
  author  = {Weindel, Franziska and Heckel, Reinhard},
  journal = {arXiv preprint arXiv:2504.00613},
  year    = {2025}
}

@article{romera2024funsearch,
  title   = {Mathematical discoveries from program search with large language models},
  author  = {Romera-Paredes, Bernardino and Barekatain, Mohammadamin and Novikov, Alexander and Balog, Matej and Kumar, M. Pawan and Dupont, Emilien and Ruiz, Francisco J. R. and Ellenberg, Jordan S. and Wang, Pengming and Fawzi, Omar and Kohli, Pushmeet and Fawzi, Alhussein},
  journal = {Nature},
  volume  = {625},
  pages   = {468--475},
  year    = {2024}
}
\bibliographystyle{alpha}

\appendix\label{app:start}
\newpage
\section{Explicit constructions}
\label{app:constructions}

All polynomials are over $\mathbb{F}_2$.  Given a block length $m$, a divisor $h$
of $x^m-1$ of degree $k$, $g=(x^m-1)/h$ and polynomials $f_1=1,f_2,\dots,f_\ell$,
the words
\[
  \bigl(bgf_1,\,bgf_2,\,\dots,\,bgf_\ell\bigr)\bmod (x^m-1),\qquad
  b=\textstyle\sum_{i<k}b_ix^i,
\]
form a quasi-cyclic code of length $m\ell$, index $\ell$ and dimension $k$
(coordinate $j$ of a block is the coefficient of $x^j$).  Some codes append
\emph{tail} coordinates after the blocks.  These depend only on
\[
  r=b \bmod (x^5-1)=(r_0,\dots,r_4),\qquad r_t=\sum_{i\,\equiv\, t \ (\mathrm{mod}\ 5)} b_i ;
\]
a tail coordinate equals $\langle c,r\rangle=\sum_t c_tr_t$ for a pattern $c$,
printed as the bit string $c_0c_1c_2c_3c_4$.  In the generator matrix whose row $j$
is the codeword of $b=x^j$ ($0\le j<k$), its entry in row $j$ is $c_{j \bmod 5}$.
The following seven base codes give all improved entries of
Table~\ref{tab:records}.

\paragraph{\boldmath$[172,21,66]$}
\begin{itemize}
  \item $m=85$, index 2.
  \item $h$ is the product of
  \begin{itemize}
    \item $x + 1$;
    \item $x^4 + x^3 + x^2 + x + 1$;
    \item $x^8 + x^6 + x^5 + x^4 + x^3 + x + 1$;
    \item $x^8 + x^7 + x^5 + x + 1$.
  \end{itemize}
  \item $f_1=1$ and $f_2=x^{13} + x^{11} + x^9 + x^6 + x^4 + x^3 + x^2 + x + 1$.
  \item Tail: $(r_0{+}\dots{+}r_4,\ r_0{+}\dots{+}r_4)$, i.e.\ the value
  $b(1)=b \bmod (x+1)$ repeated twice.  This is Construction~X with the
  repetition code $[2,1,2]$.
\end{itemize}
\paragraph{\boldmath$[173,20,68]$}
\begin{itemize}
  \item $m=85$, index 2.
  \item $h$ is the product of
  \begin{itemize}
    \item $x^4 + x^3 + x^2 + x + 1$;
    \item $x^8 + x^5 + x^4 + x^3 + x^2 + x + 1$;
    \item $x^8 + x^7 + x^6 + x^5 + x^4 + x^3 + 1$.
  \end{itemize}
  \item $f_1=1$ and $f_2=x^{15} + x^{11} + x^6 + x^3 + x^2 + 1$.
  \item Tail: 3 coordinates $\langle c_i,r\rangle$, where $c_1,c_2,c_3$ are
    \texttt{01001}, \texttt{10100}, \texttt{01010}.
\end{itemize}
\paragraph{\boldmath$[176,21,68]$}
\begin{itemize}
  \item $m=85$, index 2.
  \item $h$ is the product of
  \begin{itemize}
    \item $x + 1$;
    \item $x^4 + x^3 + x^2 + x + 1$;
    \item $x^8 + x^6 + x^5 + x^4 + x^3 + x + 1$;
    \item $x^8 + x^7 + x^5 + x + 1$.
  \end{itemize}
  \item $f_1=1$ and $f_2=x^{13} + x^{11} + x^9 + x^6 + x^4 + x^3 + x^2 + x + 1$.
  \item Tail: 6 coordinates $\langle c_i,r\rangle$, where $c_1,\dots,c_{6}$ are
    \texttt{01111}, \texttt{11111}, \texttt{11111}, \texttt{10010}, \texttt{11011}, \texttt{00101}.
\end{itemize}
\paragraph{\boldmath$[181,21,70]$}
\begin{itemize}
  \item $m=85$, index 2.
  \item $h$ is the product of
  \begin{itemize}
    \item $x + 1$;
    \item $x^4 + x^3 + x^2 + x + 1$;
    \item $x^8 + x^5 + x^4 + x^3 + x^2 + x + 1$;
    \item $x^8 + x^7 + x^3 + x + 1$.
  \end{itemize}
  \item $f_1=1$ and $f_2=x^{13} + x^{10} + x^7 + x^6 + x^4 + x^3 + 1$.
  \item Tail: 11 coordinates $\langle c_i,r\rangle$, where $c_1,\dots,c_{11}$ are
    \texttt{11000}, \texttt{10100}, \texttt{11100}, \texttt{01010}, \texttt{01110}, \texttt{10001}, \texttt{11101}, \texttt{00011}, \texttt{10011}, \texttt{00111}, \texttt{01111}.
\end{itemize}
\paragraph{\boldmath$[184,21,72]$}
\begin{itemize}
  \item $m=85$, index 2.
  \item $h$ is the product of
  \begin{itemize}
    \item $x + 1$;
    \item $x^4 + x^3 + x^2 + x + 1$;
    \item $x^8 + x^6 + x^5 + x^4 + x^2 + x + 1$;
    \item $x^8 + x^7 + x^5 + x^4 + x^3 + x^2 + 1$.
  \end{itemize}
  \item $f_1=1$ and $f_2=x^{13} + x^9 + x^8 + x^7 + x^6 + x^5 + x^3 + x + 1$.
  \item Tail: 14 coordinates $\langle c_i,r\rangle$, where $c_1,\dots,c_{14}$ are
    \texttt{11001}, \texttt{01001}, \texttt{00101}, \texttt{11101}, \texttt{11110}, \texttt{00111}, \texttt{11100}, \texttt{01010}, \texttt{10010}, \texttt{10001}, \texttt{11111}, \texttt{01110}, \texttt{10011}, \texttt{00110}.
\end{itemize}
\paragraph{\boldmath$[189,22,72]$}
\begin{itemize}
  \item $m=63$, index 3.
  \item $h$ is the product of
  \begin{itemize}
    \item $x + 1$;
    \item $x^3 + x^2 + 1$;
    \item $x^6 + x^5 + 1$;
    \item $x^6 + x^5 + x^3 + x^2 + 1$;
    \item $x^6 + x^5 + x^4 + x^2 + 1$.
  \end{itemize}
  \item $f_1=1$, $f_2=x^{14} + x^{13} + x^{12} + x^9 + x^7 + x^5 + x^3 + x^2 + 1$ and $f_3=x^{16} + x^{14} + x^{13} + x^{10} + x^7 + x + 1$.
  \item No tail.
\end{itemize}
\paragraph{\boldmath$[200,21,77]$}
\begin{itemize}
  \item Generator matrix.
  \item $A_{77}=25$, against $A_{77}=75$ for the $[200,21,77]$ code obtained by
  direct prompting.
\end{itemize}

\paragraph{\boldmath$[50,25,10]$ (double circulant)}
\begin{itemize}
  \item $m=25$, index 2.
  \item $h=x^{25}-1$, so $g=1$.
  \item $f_1=1$ and $f_2=x^{23} + x^{22} + x^{21} + x^{20} + x^{19} + x^{18} + x^{16} + x^{15} + x^{14} + x^{12} + x^8 + x^7 + x^6 + x^4 + x^2 + x$.
  \item Equivalently, the generator matrix is $[\,I_{25}\mid A\,]$, where $A$ is the
  circulant matrix whose first row is the coefficient vector of $f_2$.
  \item $d=10$, the best known minimum distance for $[50,25]$, and $A_{10}=125$.
\end{itemize}
\newpage
\subsection*{All improved entries.}  
\begin{table}[h]
  \centering\small
  \label{tab:records}
  \begin{tabular}{rrrrrl}
    \toprule
    $n$ & $k$ & $d_{\mathrm{old}}$ & $d^{\mathrm{ub}}$ & $A_d$ \\
    \midrule
    172 & 20 & 66 & 67 & 272 \\
    \textbf{173} & 20 & 67 & \textbf{68} & 1156 \\
    171 & 21 & 64 & 65 & 2040 \\
    \textbf{172} & 21 & 64 & \textbf{66} & 2720 \\
    173 & 21 & 64 & 66 & 680 \\
    174 & 21 & 65 & 66 & 680 \\
    175 & 21 & 66 & 67 & 1088 \\
    \textbf{176} & 21 & 66 & \textbf{68} & 2992\\
    177 & 21 & 66 & 68 & 2720\\
    178 & 21 & 67 & 68 & 1054 \\
    180 & 21 & 68 & 69 & 1428\\
    \textbf{181} & 21 & 68 & \textbf{70} & 3502\\
    182 & 21 & 68 & 70 & 2771 \\
    183 & 21 & 68 & 71 & 2108 \\
    \textbf{184} & 21 & 69 & \textbf{72} & 9826 \\
    185 & 21 & 70 & 72 & 8670 \\
    186 & 21 & 70 & 72 & 6188 \\
    187 & 21 & 71 & 72 & 6103 \\
    \textbf{200} & 21 & 76 & \textbf{77} & 25 \\
    201 & 21 & 77 & 78 & 264 \\
    188 & 22 & 70 & 71 & 924 \\
    \textbf{189} & 22 & 71 & \textbf{72} & 2394 \\
    \bottomrule
  \end{tabular}
  \caption{Improved entries, with the best previously known minimum distance (\emph{old}),
the minimum distance of our code (\emph{new}), and the number $A_d$ of
minimum-weight codewords of the released code.}
\end{table}

\raggedbottom

\end{document}